\documentclass [12pt]{article}

\begin{document}
\begin{center}
\textbf{INFORMATION, EXPANSIVE NONDECELERATIVE UNIVERSE AND SUPERSTRING 
THEORY}
\end{center}

\bigskip
\begin{center}
\textbf{Miroslav S\'uken\'{\i}k and Jozef \v{S}ima}

\medskip

\textbf{Slovak Technical University, Bratislava, Slovakia}
  
\end{center}

\bigskip

\textbf{Abstract}. Stemming from relationships between a number of 
information describing a system and entropy content of the system it is 
possible to determine maximal cosmological time. The contribution manifests 
a compatibility of the superstring theory and the model of Expansive 
Nondecelerative Universe.

\subsection*{Entropy and Information}

In spite of several justified objections raised to the second law of 
thermodynamics in specific cases [1], its validity for the Universe and the 
importance of entropy conception is generally accepted. Along with its 
probability meaning expressed by relation [2]
\begin{equation}
\label{eq1}
S = - k.\ln p = k.\ln P
\end{equation}
where $S$ is the entropy, $k$ is the Boltzmann constant, $p$ is the 
probability of the existence of a microstate, $P$ is the number of 
microstates associated with a macroscopic state, entropy is also a measure 
of information needed to describe system properties. In accordance with a 
common usage, entropy will be expressed as dimensionless quantity, number of 
information is given in bit. Since a number of information transferred $I$ 
by binary characters is [2, 3]
\begin{equation}
\label{eq2}
I = - \log _{2} p
\end{equation}
the entropy of a state fully describle by a number of information $I$ is 
then
\begin{equation}
\label{eq3}
S = I(k\ln 2) \cong 9\times 10^{ - 24}I
\end{equation}
Thus, a number of information needed to fully describe a system can be 
expressed either via its entropy or by a number of bits.

Applying the above ideas to the Universe with a permanent increase of its 
entropy content, it is obvious that to describe such a universe, a number of 
information must growth as well. 

\subsection*{Information, ENU and the Universe Dimensionality}

One of the corner-stones of the holographic model of the Universe, 
elaborated by Bohm [4], declares that every point of the Universe (every 
elemental particle) is in mutual contact with the other points (particles) 
and holds the information about the whole Universe. The higher 
dimensionality of the Universe, the higher number of information is needed 
to its description. A postulate relating a maximum information $I_{\max}  $ 
and a certain space dimensionality $n$ of the Universe may be generally 
expressed as
\begin{equation}
\label{eq4}
\log _{2} I_{\max}  = 2^{(n - 1)}
\end{equation}
As a starting point (rationalized latter) let us suppose that 
\begin{equation}
\label{eq5}
n = 10
\end{equation}
Then it follows from (\ref{eq4}) and (\ref{eq5})
\begin{equation}
\label{eq6}
I_{\max}  \cong 10^{154}
\end{equation}
Corresponding maximum entropy based on (\ref{eq3}) and (\ref{eq6}) is
\begin{equation}
\label{eq7}
S_{\max}  \cong 10^{131}
\end{equation}
In the ENU model [5] it holds
\begin{equation}
\label{eq:8a}
a = c.t_{c} = {\frac{{2G.m_{U}} }{{c^{2}}}} \cong 1.3\times 10^{26} \mathrm{m}
\end{equation}
where $a$ is the gauge factor, $t_{c} $ is the cosmological time, and $m_{U} 
$ is the mass of the Universe. The elementary particles did not exist at the 
very beginning of the Universe, the gravitational field quanta, however, 
existed. This is why the entropy of the Universe can be expressed by means 
of a number of gravitational field quanta at a given cosmological time. If 
the mean energy of a gravitational field quantum is denoted as $E_{g} $, 
then
\begin{equation}
\label{eq8}
S = {\frac{{m_{U} .c^{2}}}{{{\left| {E_{g}}  \right|}}}} = {\frac{{t_{c} 
.c^{5}}}{{2G.{\left| {E_{g}}  \right|}}}}
\end{equation}
As shown in our previous paper [6], it holds in the ENU
\begin{equation}
\label{eq9}
i.\hbar {\frac{{d\Psi _{g}} }{{dt}}} = E_{g} .\Psi _{g} 
\end{equation}
where $\Psi _{g} $ is the wave function of the Universe defined as
\begin{equation}
\label{eq10}
\Psi _{g} = e^{ - i(t_{Pc} .t_{c} )^{ - 1 / 2}.t}
\end{equation}
It follows from (\ref{eq9}) and (\ref{eq10}) that 
\begin{equation}
\label{eq11}
{\left| {E_{g}}  \right|} = \hbar (t_{Pc} .t_{c} )^{ - 1 / 2}
\end{equation}
where $t_{Pc} $ is the Planck time
\begin{equation}
\label{eq:13a}
t_{Pc} = \left( {{\frac{{G.\hbar} }{{c^{5}}}}} \right)^{1 / 2} = 
5.39056\times 10^{ - 44} \mathrm{s}
\end{equation}
The entropy content at a time $t$ is then, based on (\ref{eq8}),
(\ref{eq9}), and (\ref{eq10}), given by
\begin{equation}
\label{eq12}
S = \left( {{\frac{{t}}{{t_{Pc}} }}} \right)^{3 / 2}
\end{equation}
i.e. at the time being 
\begin{equation}
\label{eq13}
S \cong 10^{92}
\end{equation}
A crucial conclusion stems from (\ref{eq7}) and (\ref{eq12}). It
determines a cosmological time in which the maximum entropy should be
reached
\begin{equation}
\label{eq:16a}
t_{c(\max )} \cong 10^{35} - 10^{36} \mathrm{years}
\end{equation}
A significance of this result lies in the fact that $t_{c(\max )}$
represents the time for which a decay of of baryonic matter is
anticipated.

\subsection*{Concluding Remarks}
Standard cosmological theories deals with the 3-dimensional space. Within 
the Kaluza-Klein theoretical approaches [7] more-dimensional spaces ($n$ = 4 
to 6) are elaborated [8]. A classic superstring theory used space with $n$ = 
9, in the M-theory $n$ = 10. There are other conceptions within the 
superstring theories [9, 10] in which a number of dimensions is even 
higher. 

The ENU model is compatible with superstring theory having $n$ = 10. 
Calculations based on relations present in this work lead to a result 
showing that if $n< 10$, the maximum cosmological time would be less than 
the present time. 

New results might be obtained when some features of the ENU model (such as 
the matter creation and gravitational energy localization) are incorporated 
into superstring theory. 

\subsection*{References}

\begin{enumerate}

\item V. \v{C}\'apek, J. Bok, Physica A 290 (2001) 379

\item W. Yourgrau, A. van der Merwe, Entropy, Information and Statistical 
Thermodynamics, in Cosmology, Fusion and Other Matters (F. Reines, ed.), 
Adam Hilger, London, 1972, p.241 

\item B.H. Mahan, University Chemistry (2$^{{\rm n}{\rm d}}$ ed.),
  Addison-Wesley Publ. Comp., London, 1972, p. 340

\item D. Bohm, B.J. Hiley, The Undivided Universe, Routledge, London, 1993

\item V. Skalsk\'y, M. S\'uken\'{\i}k, Astrophys. Space Sci., 178 (1991) 169

\item M. S\'uken\'{\i}k, J. \v{S}ima, Preprint gr-qc/0010061

\item T. Appelquist, A. Chodos, P.G. Freund (Eds.) Modern Kaluza-Klein Theories, 
Addison-Wesley, Reading, MA., 1987

\item T. Gherghetta, M. Shaposhnikov, Phys. Rev. Lett., 85 (2000) 240

\item M.B. Green, J.H. Schwarz, E. Witten, Superstring Theory, Vol. 1, 2. 
Cambridge University Press, Cambridge, 1989 

\item M. Kaku, Introduction to Superstrings, Springer-Verlag, New York, 1988
\end{enumerate}

\end{document}